\documentclass[reprint,amsmath,amssymb, aps, prl]{revtex4-1}

\usepackage{graphicx}
\usepackage{dcolumn}
\usepackage{bm}
\usepackage{hyperref}
\usepackage{color}
\usepackage{gensymb}

\def\be{\begin{equation}}
\def\ee{\end{equation}}
\def\bea{\begin{eqnarray}}
\def\eea{\end{eqnarray}}

\newcommand{\levicivita}{}
\def\levicivita#1#{\tensor#1{\epsilon}}

\begin{document}
	
	
	\title{A study on black hole shadows in asymptotically de Sitter spacetimes}
	
	\author{Rittick Roy$^1$}
	\email{rittickrr@gmail.com}
	\author{Sayan Chakrabarti$^2$}%
	\email{sayan.chakrabarti@iitg.ac.in}
	\affiliation{%
		$^1$Department of Physics, Indian Institute of Technology Bombay, Mumbai 400076, India 
	}%
	\affiliation{%
		$^2$Department of Physics, Indian Institute of Technology Guwahati, Guwahati 781039, India 
	}%

	\date{\today}
	
	\begin{abstract}
		Previous works on black hole shadows have been primarily focused on studying shadows in asymptotically flat and anti-de Sitter space times. In the present work, we find general expressions for asymptotically de Sitter black hole shadow as seen by static and comoving observers, for any spherically symmetric black hole solution, in any space time dimension in generic theories of gravity. 
		As test cases, we use the derived general expressions to study the shadow of three different black hole solutions: the four dimensional Reissner-Nordstr\"{o}m de Sitter black hole, the recently proposed four dimensional Gauss Bonnet de Sitter black hole and the five dimensional Gauss Bonnet de Sitter black hole. 
	\end{abstract}
	
	\maketitle
	
	\section{Introduction}
	
	Black holes are one of the simplest objects in General Relativity, they are described by only a few parameters. Yet, they are fascinating and {deserve} attention on their own merit simply because they shed lights into the classical and quantum nature of spacetime. Although there were {many} seminal works to understand black holes theoretically, observational {evidence} {to prove their existence has been scarce}. Very recently, with the detection of gravitational waves from the merger of two black holes \cite{Abbott:2016blz, TheLIGOScientific:2016wfe, Abbott:2016nmj}, we have obtained the first evidence of the existence of black holes. Apart from these, evidences of existence of supermassive central objects at the centre of many galaxies, including our own, have come from various astrophysical data \cite{Schodel:2002vg, Ghez:2008ms}, pointing towards the existence of black holes too.    Still, all these evidences for existence of black holes can be thought of as indirect evidences, in the sense that one is not really `seeing' {(spatially resolving)}  the object in the sky. Therefore, a direct observational evidence of the existence of black holes {is} to observe its shadow as was proposed in \cite{Bardeen, Luminet:1979nyg} for the ultimate probe of its photon sphere and {has recently been  performed by the Event Horizon Telescope (EHT) collaboration \cite{EHT, Akiyama:2019brx, Akiyama:2019sww, Akiyama:2019bqs, Akiyama:2019fyp, Akiyama:2019eap}}. 
	
	The shadow of a black hole is nothing but an image of the photon sphere which is gravitationally lensed by the presence of extremely strong gravitational field around the black hole and {projected on the local sky of an observer}. In other words, any compact object can cast a shadow showing its optical appearance because of the strong gravitational field created by the compact object itself. The strong gravitational field affects everything including photons. It bends the light paths and thereby acts as a lens and distorts the image. Photons from any source behind the compact object can cast a shadow on a plane to be seen by an asymptotic observer at infinity. It has been found out that the shadows of spherically symmetric black holes are in general circular and that of rotating black holes are not precisely circular but are deformed in shape \cite{Luminet:1979nyg, Synge, Falcke:1999pj}. The variety of shadows from different {black hole spacetimes} are found to be dependent on the spin parameter (in case of rotating black holes), the configuration of the light emission region near the black hole and on the inclination angle of a black hole \cite{Hioki:2009na}. {In} \cite{Hioki:2009na}, the authors introduced two new observables characterising the apparent shape of the shadow and showed that the spin parameter and inclination angle of a Kerr black hole can be extracted by observing the shadow. Afterwards, there was a renewed interest in studying shadows of black holes and compact objects in different theories of gravity along with black holes {arising in} Einstein's General Relativity, in a widely varied manner and from different perspectives and in different spacetime dimensions. {Studies in different directions such as: apparent shapes of black hole shadows with different black hole configurations and in different spacetime geometries \cite{Bambi:2008jg, Hioki:2008zw, Virbhadra:2008ws, Johannsen:2015qca, Bozza:2006nm}, shadows of non-rotating and rotating black hole spacetimes in different modified theories of gravity \cite{Das, Konoplya:2019fpy,Konoplya:2020bxa, Kraniotis:2010gx, Amarilla:2010zq, Amarilla:2011fx, Abdujabbarov:2012bn, Amarilla:2013sj, Grenzebach:2014fha, Atamurotov:2013sca, Papnoi:2014aaa, Shaikh:2018kfv}, constraining charge of a black hole from the study of shadows \cite{Zakharov:2014lqa}, black hole shadows in dynamically evolving spacetimes \cite{Mishra}, shadows of rotating regular black holes\cite{Amir:2016cen}, proposing new methods of calculating black hole shadows \cite{Younsi:2016azx}, shadows of quantum corrected black holes and their relation to quasi-normal modes \cite{Konoplya:2019xmn, Konoplya:2019sns, Liu:2020ola} were performed}.
	
	Recent imaging of the black hole horizon at the centre of the $M87^*$ galaxy by the Event Horizon Telescope (EHT) collaboration \cite{EHT} has opened up the possibility {to probe compact objects, search for new phenomena and learn more about their
		nature.} In a series of papers \cite{EHT, Akiyama:2019brx, Akiyama:2019sww, Akiyama:2019bqs, Akiyama:2019fyp, Akiyama:2019eap}, the collaboration has described the instrumentation, data processing, imaging techniques, {theoretical interpretation} and parameter estimation process for the central massive black hole. The EHT is a large telescopic array consisting of a network of radio telescopes throughout the world. By combining a huge set of data from several very long baseline interferometry stations from all over the world, it is possible to achieve angular resolution which enables one to observe objects of the size of the event horizon of a supermassive black hole. This pathbreaking discovery has naturally attracted a lot of interest in the area of study of shadows of compact objects {recently}. A number of {studies (\cite{Bambi:2008jg, Hioki:2008zw, Virbhadra:2008ws, Johannsen:2015qca, Bozza:2006nm, Kraniotis:2010gx, Amarilla:2010zq, Amarilla:2011fx, Abdujabbarov:2012bn, Amarilla:2013sj, Zakharov:2014lqa, Grenzebach:2014fha, Atamurotov:2013sca, Amir:2016cen, Mishra, Papnoi:2014aaa, Das, Konoplya:2019fpy, Younsi:2016azx, Konoplya:2020bxa, Konoplya:2019xmn, Konoplya:2019sns,  Liu:2020ola, Shaikh:2018kfv})} were performed already before the construction of the first image for a wide class of spacetimes. The flurry of work began after the construction of the first image and within a short span of time several interesting and fruitful papers {appeared} in the literature, {focusing on different features of black hole shadows  \cite{Shaikh:2019fpu,Igata:2019pgb, Cunha:2017qtt}, observational signatures of exotic compact objects \cite{Bhattacharya:2017chr, Gyulchev:2019tvk,Shaikh:2019hbm}, tests of General Relativity using the shadow of black holes \cite{Glampedakis:2018blj, Psaltis:2018xkc,Vagnozzi:2019apd, Abdikamalov:2019ztb,Tian:2019yhn,  Bambi:2019tjh}, spin measurements of rotating black holes using shadows \cite{Tamburini:2019vrf, Wei:2019pjf}, black hole shadows in different modified theories of gravity \cite{Moffat:2019uxp,Allahyari:2019jqz,cunha}, EHT constraint on the ultralight scalar hairs \cite{Davoudiasl:2019nlo, Roy:2019esk,  Jusufi:2019nrn,   Cunha:2015yba, Cunha:2019ikd}, modifications of spinning and non-spinning black hole spacetimes inspired by asymptotically safe quantum gravity and its effect on black hole shadow \cite{Held:2019xde} }. Although the particular area of study on black hole shadows includes all different asymptotic structures of spacetime, surprisingly we found that there has not been much work related to black holes in asymptotically de Sitter spacetime {except for a few  \cite{Perlick, Li:2020drn, Stuchlik:2018qyz}}. Since our universe is known to be expanding \cite{Carroll:2000fy, Aghanim:2018eyx, Riess:1998cb, Perlmutter:1998np}, it is important to study the asymptotically de Sitter spacetime more closely and see how the shadow of a black hole would change with the cosmic expansion. An interesting study in this respect, for the Schwarzschild de Sitter spacetime, has been recently performed in \cite{Perlick} (and a {work on shadows} in the Kerr de Sitter spacetime in \cite{Li:2020drn} and light escape cones in certain special class of reference frames in \cite{Stuchlik:2018qyz}). In this work,  the authors {consider} two different classes of observers: static observers, whose spatial position is fixed and observers comoving with the cosmic expansion. {The shadow with respect to the comoving observer has been found by performing a coordinate transformation to go from the asymptotically de Sitter black hole metric to a form of expanding universe metric with a black hole embedded in it. This type of transformation was first studied in \cite{McVittie} for the Schwarzschild de Sitter spacetime. We would be generally referring to such transformations as McVittie type transformation, from here onwards.} As a first step, the above transformation requires expressing the black hole metric in an {isotropic} form. But it is not always possible to find a closed form solution for any general black hole metric in an isotropic form, let alone in a McVittie type coordinate system. {This raises the question}: is it possible to find an expression for the shadow with respect to a comoving observer for black hole metric without a closed form solution for the isotropic transformation function? Interestingly, we found that one can indeed find the comoving shadow even without knowing the exact transformation functions as mentioned above. 
	
	There has always been interests in black holes in modified theories of gravity (see \cite{Clifton:2011jh, Nojiri} for reviews) and in particular in theories containing higher curvature corrections to Einstein-Hilbert action, {e.g. in} several developments in string theory \cite{Zwiebach:1985uq}. In this work, along with one black hole background (Reissner-Nordstr\"{om} de Sitter) arising out of Einstein's General Relativity, we also focus on black holes in Gauss Bonnet gravity in asymptotically de Sitter spacetime. With the recent discovery of a four dimensional Gauss Bonnet de Sitter (GB dS) \cite{Glavan:2019inb} solution, we focus on the same, along with the shadow of GB dS black hole in higher dimensions (in this paper we focus on $d=5$ only) too. 
	
	The paper is structured as follows: we start with a discussion of the velocity of the comoving observer with respect to the static observer and find a general expression for the velocity in Section I following the methodology prescribed by \cite{Perlick}, but extending it to arbitrary spherically symmetric black holes. Using the results of Section I, we find the expressions for the static and comoving shadow in a generic four dimensional spherically symmetric black hole in de Sitter spacetime. In Section II and apply this to study the shadow of the Reissner-Nordstr\"{o}m-de Sitter (RN-dS) black hole and the four dimensional Gauss Bonnet de Sitter black hole spacetime. In Section III, we find a general expression for shadow in higher dimensions (we confine to five dimensions only for this purpose, but it can in principle be generalised to $d>5$) and find the shadow of the Gauss Bonnet de Sitter black hole spacetime. We conclude our work in Section IV. Throughout the rest of the work we will be using $G=c=1$ system of units, unless mentioned otherwise. 
	\section{Velocity of comoving observers with respect to a static observer}
	We {consider} a spherically symmetric spacetime describing an asymptotically de Sitter black hole solution in $d$-spacetime dimensions. The metric {is} written in a general form as 
	\begin{equation}\label{eq:genMetric}
	ds^2=-h(r)dt^2+f(r)dr^2+r^2d\Omega_{d-2}^2
	\end{equation}
	where $d\Omega_{d-2}^2$ is the line element of the $(d-2)$ dimensional unit sphere. It should be noted that this metric must result from the Einstein's equations in a well-defined gravity theory with a cosmological constant and for a well-defined stress-energy tensor. The functions $h(r)$ and $f(r)$ are parameterized by the black hole parameters and the cosmological constant $\Lambda$.
	
	The original de Sitter metric and the zero-curvature FLRW solution {is} related by a coordinate transformation. We would like to express the black hole de Sitter metric of Eq. (\ref{eq:genMetric}) in a McVittie type coordinate system \cite{McVittie} i.e. in the form of a black hole solution embedded in an expanding universe. The observers on constant spatial curves in this spacetime moves along with the cosmic expansion and hence defines the `comoving' class of observers. To express Eq. (\ref{eq:genMetric}) in such a coordinate system, it would be convenient to express the {asymptotically} flat black hole metric in an isotropic form and use these coordinates further, such that it is easier to relate the isotropic metric of the expanding universe and the black hole metric via the McVittie type coordinates. Following this, we define two functions, $h_0(r)$ and $f_0(r)$, such that  $h_0(r)=h(r)|_{\Lambda=0}$ and $f_0(r)=f(r)|_{\Lambda=0}$.
	The black hole solution in the asymptotically flat case therefore is given by 
	\begin{equation}\label{eq:flatMet}
	ds_{flat}^2=-h_0(r)dt^2+f_0(r)dr^2+r^2d\Omega_{d-2}^2
	\end{equation}
	We can then consider a coordinate transformation of the radial coordinate $r\rightarrow {\tilde r}$, to express the metric in Eq. (\ref{eq:flatMet}) in an isotropic form. When written in isotropic form, Eq. (\ref{eq:flatMet}) should look like
	\begin{equation}\label{eq:isoComp}
	ds^2_{flat}=-h_0(r)dt^2+g^2(\tilde r)(d\tilde r^2+\tilde r^2d\Omega_{d-2}^2)
	\end{equation}
	Note that the $g_{tt}$ component would have some other form in the new coordinate system, say $k(\tilde r)$, but we have retained it in the original coordinates since $h_0(r)=k(\tilde r)$. Comparing Eq. (\ref{eq:flatMet}) and Eq. (\ref{eq:isoComp}) we get
	\bea
	r^2&=&g^2(\tilde r) \tilde r^2, \label{eq:isoComp1}\\
	f_0(r)dr^2&=&g^2(\tilde r)d\tilde r^2
	\eea
	Therefore, we can express the differential element of $r$ in terms of the differential of $\tilde r$ as
	\begin{equation}\label{eq:drdri}
	dr=\frac{r}{\tilde r\sqrt{f_0(r)}}d\tilde r
	\end{equation}
	Note that there would a positive and negative solution to the above equation but we chose the positive solution since we physically require $\frac{dr}{d\tilde r}\geq 0$. Now expressing Eq. (\ref{eq:genMetric}) in this new coordinate system we get
	\begin{equation}\label{eq:genIso}
	ds^2=-h(r)dt^2+\frac{f(r)r^2}{f_0(r)\tilde r^2}d\tilde r^2+g^2(\tilde r)\tilde r^2d\Omega_{d-2}^2
	\end{equation}
	Following McVittie's work \cite{McVittie}, we make another coordinate transformation
	\begin{equation}\label{eq:rirc}
	\tilde r=r_ca(t_c)
	\end{equation}
	where $a(t_c)$ is the cosmic scale factor which reduces to unity in the case $\Lambda=0$. Taking the differential of Eq. (\ref{eq:rirc}) we get
	\begin{equation}\label{eq:dri}
	d\tilde r=a\Bigg(\frac{\dot a}{a}r_cdt_c+dr_c\Bigg)
	\end{equation}
	where a dot represents a derivative with respect to the $t_c$ coordinate. The transformation of the original temporal coordinate $t$ to these new coordinates $(t_c, r_c)$ is taken in some general form as 
	\begin{equation}\label{eq:dt}
	dt=B(t_c, r_c)dt_c+C(t_c, r_c)dr_c
	\end{equation}
	Expressing Eq. (\ref{eq:genIso}) in this new coordinate system we get 
	\begin{align}\label{eq:genComov}
		ds^2=-\Bigg(&hB^2-\frac{fr^2\dot a^2}{f_0 a^2}\Bigg)dt_c^2+ \Bigg(\frac{fr^2a^2}{f_0\tilde r^2}-hC^2\Bigg)dr_c^2\nonumber \\ 
		+&2\Bigg(\frac{\dot afr^2}{af_0r_c}-hBC\Bigg)dt_cdr_c +g^2\tilde r^2d\Omega_{d-2}^2 
	\end{align}
	We {now impose} two conditions on the metric in Eq. (\ref{eq:genComov}): firstly the metric should be isotropic and secondly, it should be diagonal. These two conditions are motivated by the fact that both the spherically symmetric black hole metric and the metric for the expanding universe are diagonal, and since we are using the isotropic $\tilde r$ coordinates, both metrics should also be isotropic. Requiring the condition that the metric should be isotropic, the metric functions in Eq. (\ref{eq:genComov}) should satisfy $g_{rr}$=$g_{\theta \theta}/r_c^2$. Solving this, we get
	\begin{equation}\label{eq:C}
	C(t_c, r_c)=\frac{r}{r_c \sqrt{h(r)}}\Bigg(\frac{f(r)}{f_0(r)}-1\Bigg)^{1/2}.
	\end{equation}
	Equating the off diagonal term, of the metric in Eq. (\ref{eq:genComov}), $g_{tr}$=0 and using Eq. (\ref{eq:C}), we get 
	\begin{equation}\label{eq:B}
	B(t_c, r_c)=\frac{\dot a(t_c) r f(r)}{a(t_c)f_0(r)\sqrt{h(r)}}\Bigg(\frac{f(r)}{f_0(r)}-1\Bigg)^{-1/2}.
	\end{equation}
	Therefore the transformation from the original coordinates $(t,r)$ to $(t_c, r_c)$ {is} given by using Eq. (\ref{eq:drdri}), (\ref{eq:dri}), (\ref{eq:dt}), (\ref{eq:C}) and (\ref{eq:B}) as
	\begin{align}
		dt=\frac{\dot a(t_c) r f(r)}{a(t_c)f_0(r)\sqrt{h(r)}}&\Bigg(\frac{f(r)}{f_0(r)}-1\Bigg)^{-1/2}dt_c \nonumber \\
		&+\frac{r}{r_c \sqrt{h(r)}}\Bigg(\frac{f(r)}{f_0(r)}-1\Bigg)^{1/2}dr_c
	\end{align}
	\begin{equation}
	dr=\frac{\dot a(t_c)r}{a(t_c)\sqrt{f_0(r)}}dt_c+\frac{r}{r_c\sqrt{f_0(r)}}dr_c
	\end{equation}
	With the above transformation, the metric in Eq. (\ref{eq:genComov}) has the required behaviour i.e. in the case of $\Lambda=0$, it reduces to the black hole solution and in the case of all the black hole parameters going to zero, it reduces to the form of the isotropic expanding universe solution. Now, we follow a similar approach as in \cite{Perlick} to find the comoving velocity. Writing the above in Gaussian vector form we get
	\begin{align}
		\frac{\partial}{\partial t_c}=\frac{\dot a(t_c) r f(r)}{a(t_c)f_0(r)\sqrt{h(r)}}&\Bigg(\frac{f(r)}{f_0(r)}-1\Bigg)^{-1/2}\frac{\partial}{\partial t}\nonumber \\
		&+\frac{\dot a(t_c)r}{a(t_c)\sqrt{f_0(r)}}\frac{\partial}{\partial r}
	\end{align}
	\begin{equation}
	r_c\frac{\partial}{\partial r_c}=\frac{r}{ \sqrt{h(r)}}\Bigg(\frac{f(r)}{f_0(r)}-1\Bigg)^{1/2}\frac{\partial}{\partial t}+\frac{r}{\sqrt{f_0(r)}}\frac{\partial}{\partial r}
	\end{equation}
	The velocity of a comoving observer with respect to a static observer {is} given by the special-relativistic formula 
	\begin{equation}\label{eq:srVel}
	g_{\mu\nu} U^\mu_{\rm{stat}}U^\nu_{{\rm comov}}=\frac{-1}{\sqrt{1-v^2}}
	\end{equation}
	where $U^\mu_{\rm stat}$ and $U^\nu_{\rm comov}$ are the velocity vectors of the static and the comoving observers, written in the Gaussian unit vectors form as \begin{equation}
	U^\mu_{{\rm stat}}\frac{\partial}{\partial x^\mu}=N_{\rm stat}\frac{\partial}{\partial t}
	\end{equation}
	\begin{align}
		&U^\mu_{\rm comov}\frac{\partial}{\partial x^\mu}=N_{\rm comov}\frac{\partial}{\partial t_c} \nonumber \\
		&=N_{\rm comov}\Bigg(\frac{\dot a(t_c) r f(r)}{a(t_c)f_0(r)\sqrt{h(r)}}\Bigg(\frac{f(r)}{f_0(r)}-1\Bigg)^{-1/2}\frac{\partial}{\partial t}\nonumber \\
		&+\frac{\dot a(t_c)r}{a(t_c)\sqrt{f_0(r)}}\frac{\partial}{\partial r}\Bigg),
	\end{align}
	where $N_{\rm stat}$ and $N_{\rm comov}$ are just normalisation factors determined by the normalisation condition $ g_{\mu\nu} U^\mu U^\nu =-1$, as
	\bea
	N_{\rm stat}&=&\frac{1}{\sqrt{h(r)}},\\
	N_{\rm comov}&=&\frac{a(t_c)}{\dot a(t_c)r}\Bigg(1-\frac{f_0(r)}{f(r)}\Bigg)^{1/2}.
	\eea
	Substituting {all this information} in Eq. (\ref{eq:srVel}), we get the velocity of the observer comoving with the cosmic expansion with respect to the static observer as 
	\begin{equation}\label{eq:genVel}
	v_{\rm comov}=\sqrt{1-\frac{f_0(r)}{f(r)}}
	\end{equation}
	Thus, we see that the comoving velocity does not depend explicitly on the transformation functions or the scaling function $a(t_c)$, but only depends on the metric function $f(r)$. For any general spherically symmetric black hole solution in an asymptotically de Sitter universe, we could find the velocity of the comoving observer with respect to the static observer using the above expression. 
	\section{Shadow of black holes in Four Dimensions}
	In four space time dimensions, we can write the metric Eq. (\ref{eq:genMetric}) as
	\begin{equation}\label{eq:metric4D}
	ds^2=-h(r)dt^2+f(r)dr^2+r^2d\theta^2+r^2\sin^2\theta d\phi^2.
	\end{equation}
	We have chosen to work with the forms of $h(r)$ and $f(r)$ in the above metric which describes a black hole solution in an asymptotically de Sitter universe. Hence, there should be {at least} two roots of the equation $h(r)=0$ corresponding to the position of the event horizon $r_+$ and the cosmological horizon $r_+^c$. Using the Hamilton-Jacobi formalism and exploiting the symmetries of the system, the null geodesics in this spacetime {is} described by a set of equations as
	\bea
	\dot t&=&h^{-1}(r), \label{eq:dott1}\\
	\dot \phi&=&\frac{\eta_\phi}{r^2  \sin^2\theta}, \label{eq:dotphi1}\\
	\dot r^2&+&V_{\rm eff}(r)=0, \label{eq:dotr1}\\
	r^4\dot \theta^2&=&\chi-\eta_\phi^2 /\sin^2\theta, \label{eq:dottheta1}
	\eea
	where $\chi$ and $\eta_\phi$ are two constants of motion related to the conserved energy and angular momentum of the photons and the dot represents a derivative with respect to the affine parameter. The function, $V_{eff}(r)$ is the effective potential for the radial motion and is defined as
	\begin{equation}\label{eq:veff}
	V_{\rm eff}(r)=\Bigg(\frac{\chi}{r^2f(r)}-\frac{1}{f(r)h(r)}\Bigg).
	\end{equation}
	The shadow contour of a black hole is formed by the unstable orbiting photons around the black hole, for which $V_{\rm eff}(r_p)=0$, $V_{\rm eff}'(r_p)=0$ and $V_{\rm eff}''(r_p)<0$, with $r_p$ being the position of the unstable photon orbit or the radius of the photon sphere. Solving $V_{\rm eff}(r_p)=0$ for $r_p$, we get the radius of the photon sphere. {Black hole spacetimes are known to feature photon spheres} and hence it is always possible to find at least one real positive root of the above equation outside the event horizon. Substituting this value in the second condition,  $V_{\rm eff}'(r_p)=0$, we solve for the constant of motion, $\chi$, for the unstable orbiting photons. 
	
	Now consider an observer, fixed at some spatial position $O$ at the equatorial plane $(r_ o, \theta_o=\pi/2, \phi_o)$ such that $r_p\leq r_o\leq r_+^c$. Note that the observer must be located outside the photon sphere for the shadow contour to be visible. Since the system is spherically symmetric, we could consider the photon geodesics in the equatorial plane without any loss of generality. The shadow contour of the black hole in the local sky of the observer {is formed} by the unstable photon orbits. The angular radius $\Omega$ of the shadow contour {is given by}
	\begin{equation}\label{eq:angdia4D}
	\tan \Omega = \lim_{\Delta x\to0}\frac{\Delta y}{\Delta x}
	\end{equation}
	where $(x,y)$ are coordinates directed towards the radial and azimuthal direction in the locality of the observer (see Fig. 3 from \cite{Perlick}). From the metric in Eq. (\ref{eq:metric4D}), we can therefore write this in the form
	\begin{equation}\label{eq:angdia4D2}
	\tan \Omega = \frac{r_o\sin\theta_o}{\sqrt{f(r_o)}} \Bigg(\frac{d\phi}{dr}\Bigg)_o
	\end{equation}
	Squaring both side of Eq. (\ref{eq:angdia4D2}) and using the geodesic equations, Eq. (\ref{eq:dotphi1}) and Eq. (\ref{eq:dotr1}), to obtain $(\frac{d\phi}{dr})$ we get
	\begin{equation}\label{eq:angdia2}
	\tan^2 \Omega = \frac{-\chi_p}{r_o^2f(r_o) V_{eff}(r_o)}
	\end{equation}
	where, $\chi_p$ is the constant of motion for the unstable orbiting photons found in the previous section. Therefore, the angular radius for the shadow of a spherically symmetric black hole in four space time dimensions with respect to a static observer at $O$ is given by
	\begin{equation}\label{eq:genStatShdw}
	\sin^2\Omega=\frac{h(r_o)\chi_p}{r_o^2}
	\end{equation}
	The angular radius of the shadow with respect to the comoving and the static observer is related by the aberration formula 
	\begin{equation}
	\sin^2\Omega_{\rm comov}=(1-v^2)\frac{\sin^2\Omega}{(1-v\cos\Omega)^2}
	\end{equation}
	where, $v=v_{\rm comov}$ is the velocity of the comoving observer with respect to the static observer. Using the expression for the comoving velocity from Eq. (\ref{eq:genVel}), we get the general expression for the angular radius of the shadow of an asymptotically deSitter black hole with respect to a comoving observer as
	\begin{equation}\label{eq:genComShdw}
	\sin^2\Omega_{\rm comov}=\frac{f_0(r_o)}{f(r_o)}\frac{\chi_p h(r_o)/r_o^2}{\Bigg(1\pm\sqrt{1-\frac{f_0(r_o)}{f(r_o)}}\sqrt{1-\frac{\chi_p h(r_o)}{r_o^2}}\Bigg)^2}
	\end{equation}
	{Note that} the above formula reduces to Eq. (36) from \cite{Perlick} in case of {Schwarzschild} de Sitter metric. As an example, we apply this formula to study the shadow of the Reissner Nordstr\"{o}m de Sitter black hole and the four dimensional Gauss Bonnet de Sitter black hole in the next subsection.
	
	\subsection{Shadow of the Reissner-Nordstr\"{o}m de Sitter black hole}
	
\begin{figure}[h]
		\includegraphics[width=8.5cm]{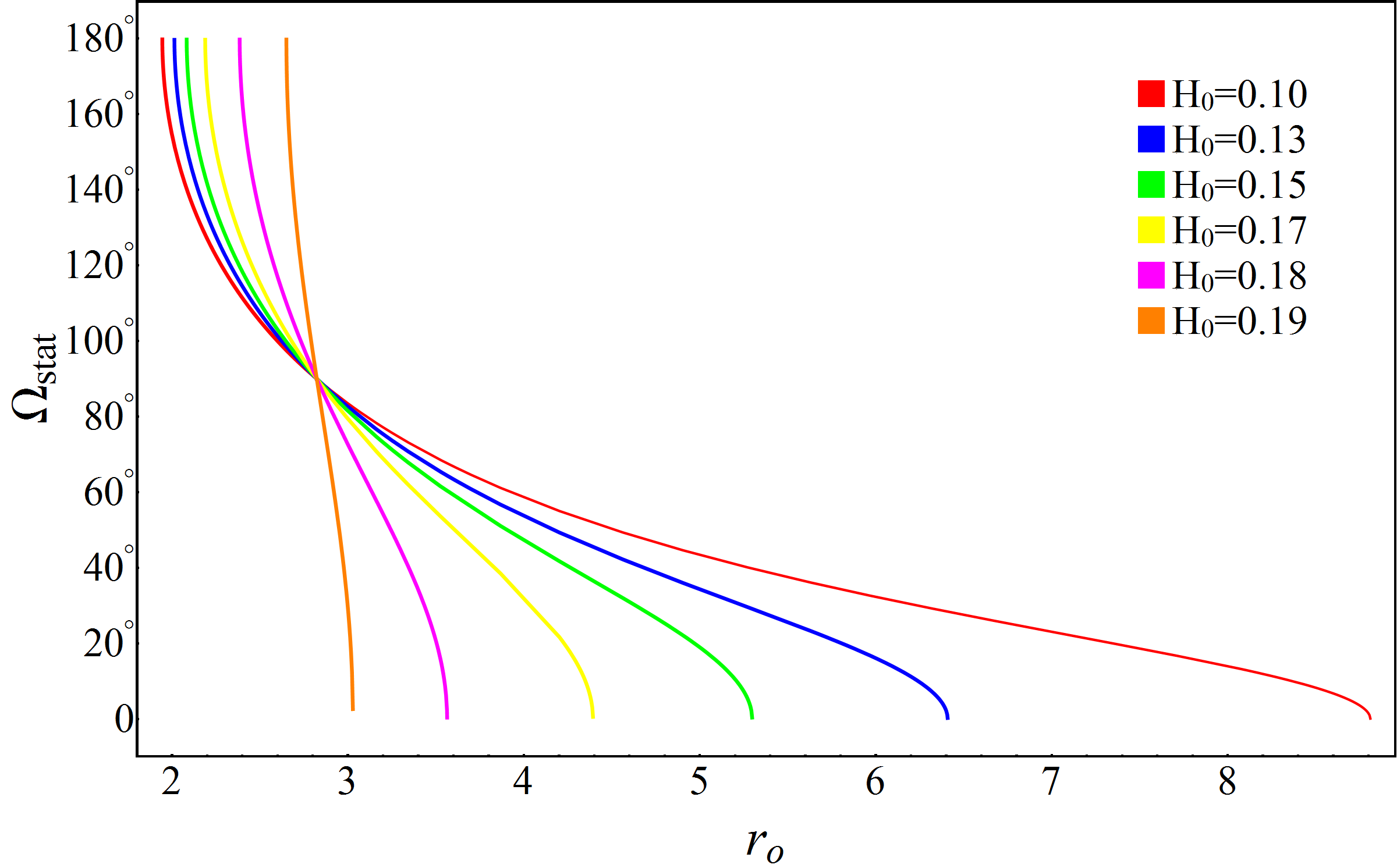}
		\centering
		\caption{The figure plots the behaviour of the static angular radius of the shadow with $r_o$ for $M=1$, $Q=0.5$ and six different values of $H_0$}
		\label{fig:RNdSH0stat}
	\end{figure}

\begin{figure}[h]
		\includegraphics[width=8.5cm]{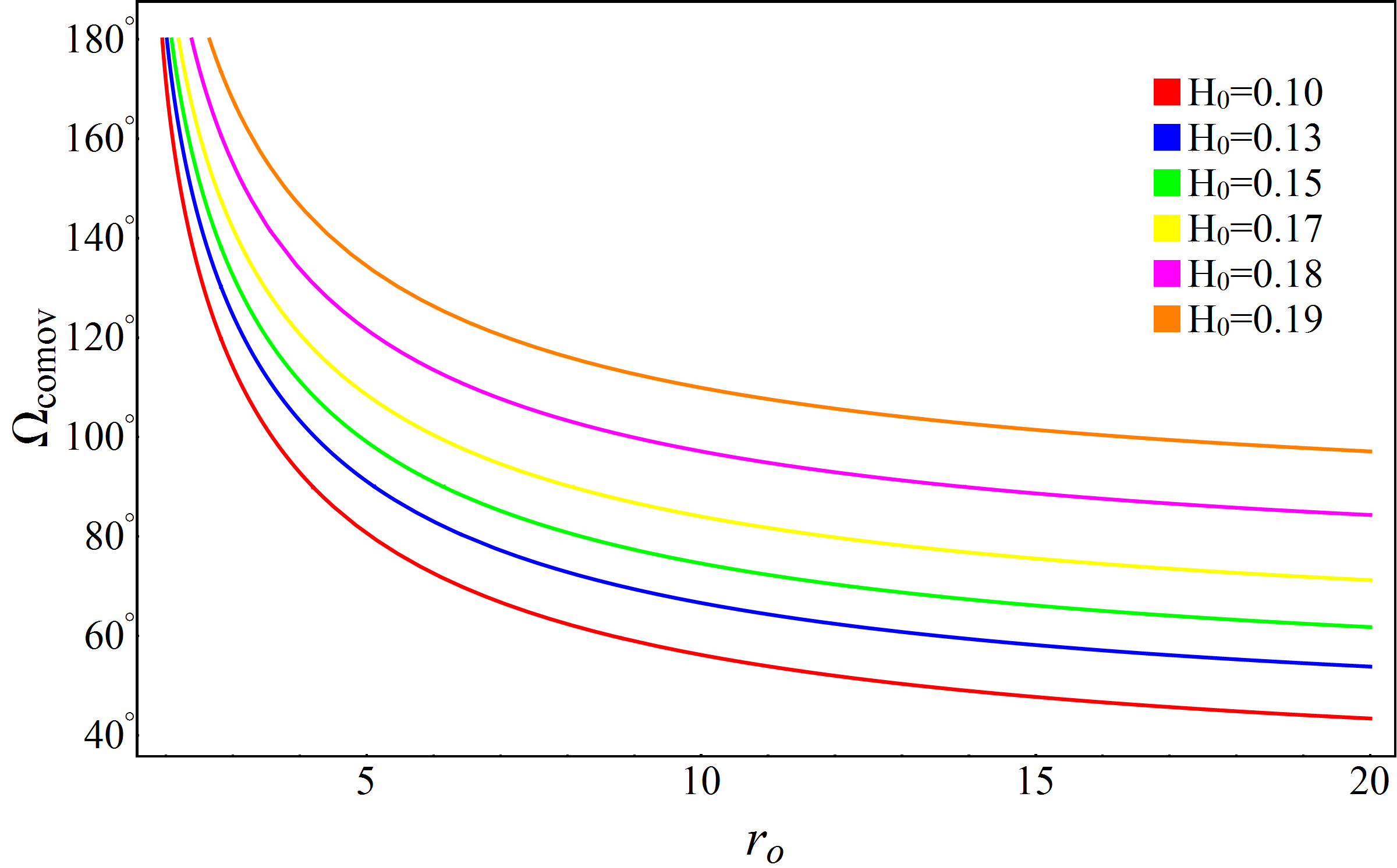}
		\centering
		\caption{The figure plots the behaviour of the comoving angular radius of the shadow with $r_o$ for $M=1$, $Q=0.5$ and six different values of $H_0$}
		\label{fig:RNdSH0}
	\end{figure}

\begin{figure}[h]
		\includegraphics[width=8.5cm]{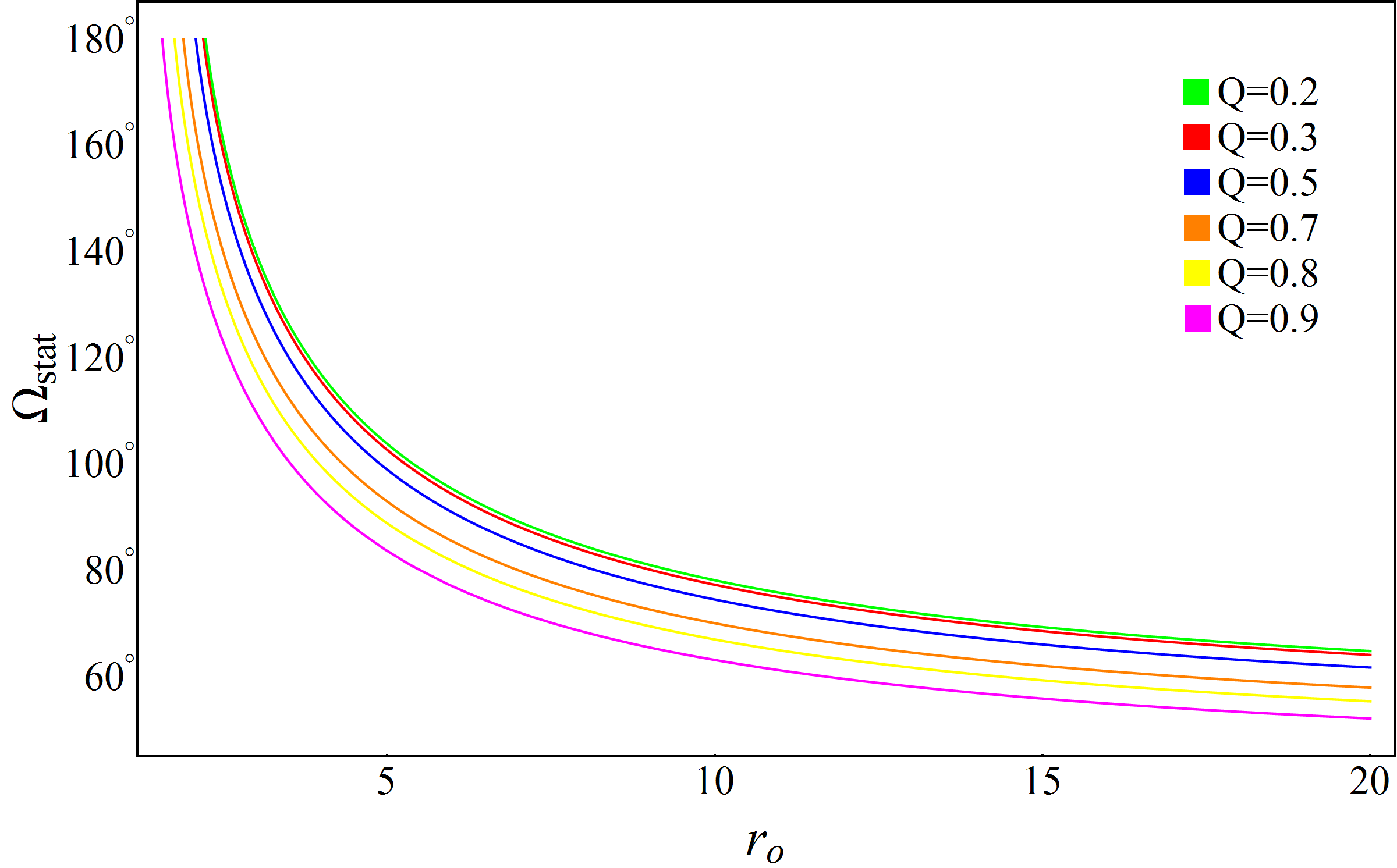}
		\centering
		\caption{The figure plots the behaviour of the comoving angular radius of the shadow with $r_o$ for $M=1$, $H_0=0.15$ and six different values of $Q$}
		\label{fig:RNdSQ3}
	\end{figure}

	The Reissner Nordstr\"{o}m de Sitter black hole spacetime is given by substituting the metric functions $h(r)$ and $f(r)$ in Eq. (\ref{eq:metric4D}) with
	\begin{equation}
	h(r)=f^{-1}(r)=1-\frac{2M}{r}+\frac{Q^2}{r^2}-\frac{\Lambda}{3}r^2
	\end{equation}
	where $M$ and $Q$ denotes the black hole mass and charge respectively. The effective potential for radial motion of photons in this spacetime is given using Eq. (\ref{eq:veff}) as
	\begin{equation}
	V_{eff}(r)=\frac{\chi h(r)}{r^2}-1
	\end{equation}
	We solve the unstable photon geodesic condition using this equation numerically to find $r_p$ and $\chi$. Fig. \ref{fig:RNdSH0stat} and Fig. \ref{fig:RNdSH0} plots the behaviour of the static and comoving angular radius with the radial position $r$ of the comoving observer for different values of {the Hubble constant}, $H_0$ (=$\sqrt{\Lambda/3}$). Fig. \ref{fig:RNdSQ3} plots the comoving shadow for different values of the black hole charge. 
	
	{The} static observers are constrained to stay between the two horizons, $r_+$ and $r_+^c$, of the spacetime. Note that the static shadow reflects this fact in Fig. \ref{fig:RNdSH0stat} by approaching $0$ at the cosmological horizon.
	On the other hand, the comoving observer can exist outside of the cosmological horizon and hence the comoving shadow is extended till spatial infinity, reaching a constant value asymptotically.
	The asymptotic behaviour of the comoving shadow in the Reissner Nordstr\"{o}m de Sitter spacetime {is} given by a simple expression:
	\begin{equation}\label{eq:asympComov}
	\lim_{r_o\rightarrow\infty} \sin^2\Omega_{comov}=\frac{1}{1+1/H_0^2\chi_p}
	\end{equation}
	The asymptotic behaviour of the static shadow could only be defined if the cosmological horizon $r_+^c$ is very far away such that the radial position of the static observer could be taken to be very large. If $r_+^c\gg M$ is satisfied, such that $r_o\gg M$ is valid, then the asymptotic shadow for a far away static observer {is} given by 
	\begin{equation}
	\lim_{r_o\gg m}\sin^2\Omega=\Bigg(\frac{1}{r_o^2}-H_0^2\Bigg)\chi_p.
	\end{equation}
	For a black hole with a fixed mass and charge, the asymptotic shadow, with respect to the comoving observer, depends only on the cosmological constant, as shown in Fig. \ref{fig:RNdSH0} and hence could be used as a probe to determine the value of $H_0$. Note that in case of the Reissner Nordstr\"{o}m de Sitter spacetime, $\chi_p$ is independent of $H_0$ and hence the entire dependence of the asymptotic shadow on the Hubble's constant comes from the $H_0^2$ factor in the denominator of Eq. (\ref{eq:asympComov}).
	\subsection{Shadow of the four dimensional charged Gauss Bonnet de Sitter black hole} 
	There has always been interests in black holes in theories containing higher curvature corrections to Einstein-Hilbert action, {motivated for example}, in string theory \cite{Zwiebach:1985uq}. It is well-known that low energy limits of string theories give rise to effective gravity models in higher dimensions and most of such models of gravity involve higher powers of the Riemann tensor in the action in addition to the usual Einstein-Hilbert term \cite{Scherk:1974ca}. In this sense, the term acts like a modification of usual Einstein's theory of {General Relativity}. Among these higher powers of Riemann curvature, the Gauss Bonnet combination is of special interest. The Gauss Bonnet term has also gained importance in the context of studying {quantum} field theory in curved space time \cite{Birell}. It is often very hard to find analytical {solutions} to {Einstein's equations} in such theories owing to the coupled non-linear nature of these equations and hence theories which does provide {analytical solutions} are of special interest. One such class of modified gravity theories is the Gauss Bonnet gravity theory. This particular theory has also gained special interest because of the fact that even if the action contains higher derivative terms, the equations of motion resulting from the action has no more than {second derivatives} of the metric and there exists exact {analytical solutions} with spherical symmetry \cite{deser, Wiltshire:1988uq}. 
	
	\begin{figure}[h]
		\includegraphics[width=8.5cm]{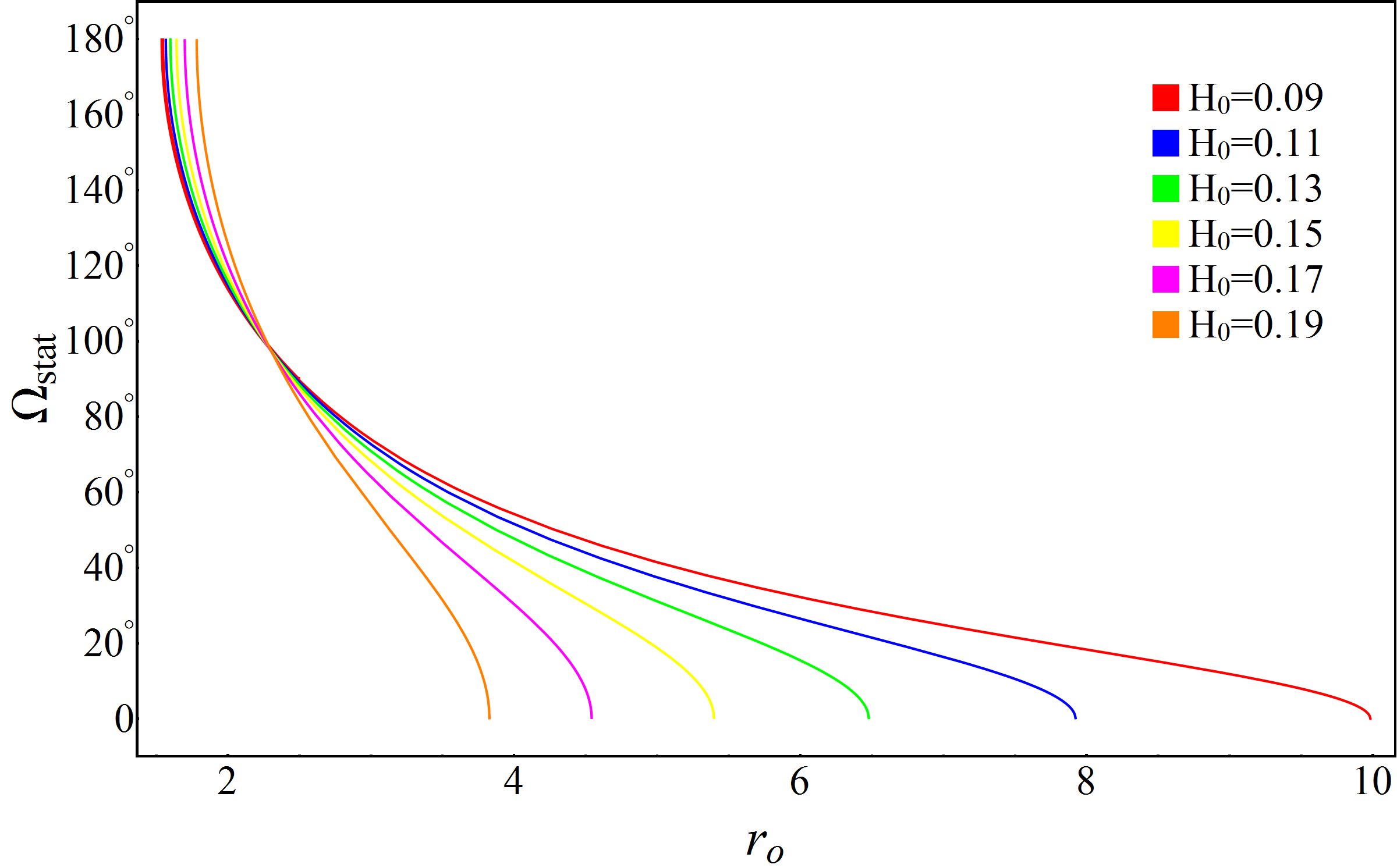}
		\centering
		\caption{The figure plots the behaviour of the comoving angular radius of the shadow of 4 dimensional charged GB dS black hole with $r_o$ for $M=1$, $Q=0.5$ and six different values of $H_0$}
		\label{fig:4DGBRNdSH0stat}
	\end{figure}
	\begin{figure}[h!]
		\includegraphics[width=8.5cm]{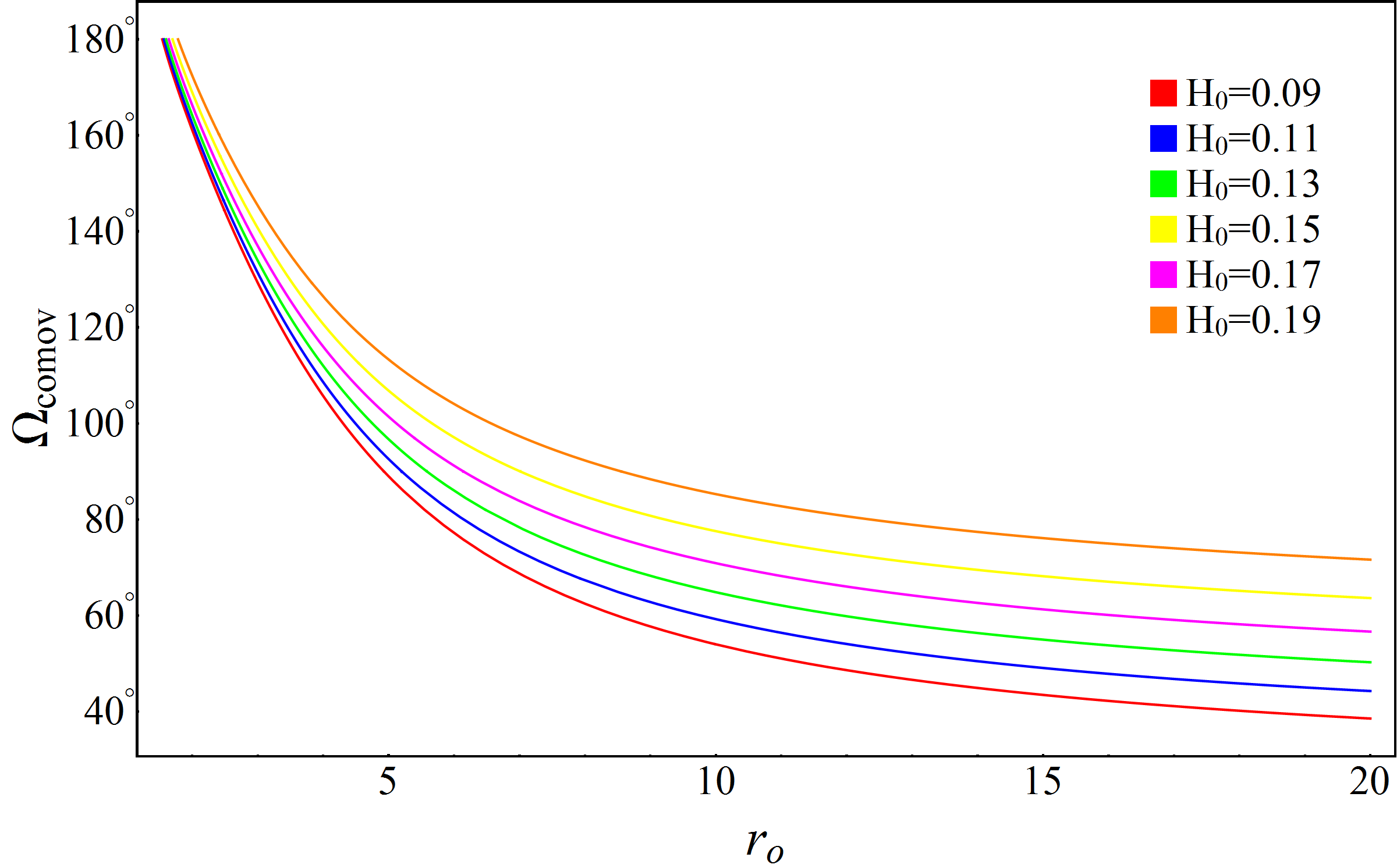}
		\centering
		\caption{The figure plots the behaviour of the comoving angular radius of the shadow of 4 dimensional charged GB dS black hole with $r_o$ for $M=1$, $Q=0.5$ and six different values of $H_0$}
		\label{fig:4DGBRNdSH0}
	\end{figure}
	
	Until recently, black holes in Gauss Bonnet gravity theories could only be defined for $d\geq 5$ dimensions (we also study the five dimensional case in the next section), since the Gauss Bonnet term
	\begin{equation}\label{eq:gb}
	\mathcal{L}_{GB}=R_{\mu \nu\gamma\delta}R^{\mu\nu\gamma\delta}-4R_{\mu\nu}R^{\mu\nu}+R^2,
	\end{equation}
	is a total derivative in $d=4$ dimension and hence it does not have any contribution to gravitational dynamics in four dimension. But recently, Glavan and Lin formulated in their study \cite{Glavan:2019inb} an Einstein-Gauss Bonnet theory in four dimension by taking a rescaled coupling constant $\alpha\rightarrow 16\pi\alpha/(d-4)$ and defining the theory as a $d\rightarrow 4$ limit of a higher dimensional theory. This theory admits a spherical charged asymptotically de Sitter black hole solution given by the metric functions \cite{Glavan:2019inb},
	\begin{equation}
	\tilde h(r)=\tilde f^{-1}(r)=1+\frac{r^2}{2\alpha}\Bigg(1\pm\sqrt{1+\frac{8\alpha M}{r^3}-\frac{4Q^2}{r^4}+4\alpha H_0^2}\Bigg)
	\end{equation}
	This {has recently attracted a lot of interest and many
		studies} regarding quasi-normal frequencies, shadows etc. \cite{Wei:2020ght,Aragon:2020qdc,Liu:2020vkh,Jin:2020emq,Yang:2020czk,Kumar:2020owy,Zeng:2020dco,Devi:2020uac}. In our study, we consider the branch with the negative sign with a positive coupling constant $\alpha$ in the above equation. We plot the static and the comoving shadow of the four dimensional Gauss Bonnet de Sitter black hole in Fig. \ref{fig:4DGBRNdSH0stat} and Fig. \ref{fig:4DGBRNdSH0} with the observer position $r_o$ for different values of the cosmological constant $H_0$. The asymptotic behaviour of the comoving and static shadow could be written in a similar form as the Reissner Nordstr\"{o}m de Sitter case
	\bea
	\lim_{r_o\rightarrow\infty} \sin^2\Omega_{\rm comov}&=&\frac{1}{1+1/H_{eff}^2\chi_p},\\
	\lim_{r_o\gg m}\sin^2\Omega&=&\Bigg(\frac{1}{r_o^2}-H_{eff}^2\Bigg)\chi_p,
	\eea
	where we define an effective Hubble's constant $H_{\rm eff}$ as
	\begin{equation}
	H_{\rm eff}^2=\frac{1}{2\alpha}(\sqrt{1+4\alpha H_0^2}-1).
	\end{equation}
	
	A noticeable difference in this case and the RN-dS case is the dependence of the asymptotic behaviour of the comoving shadow on the cosmological constant. 
	This {is} explained by looking at the above equations, whereas in case of RN-dS black hole, the asymptotic value of the shadow {depends} on $H_0^2$, here it depends on $H_{eff}^2$, which in turn depends on root of $H_0^2$.

	\section{Shadow of black holes in five dimensions}
	
	In this section we would like to focus on the method that has been employed to study shadows of higher dimensional black holes. For our purpose and for simplicity, we focus on five dimensional case, however this result is easily generalised in $d$-dimensional space time also.
	
	The general spherically symmetric spacetime in five dimensions is given by
	\begin{equation}\label{eq:genMetric5D}
	ds^2=-h(r)dt^2+f(r)dr^2+r^2d\Omega_{3}^2,
	\end{equation}
	where $d\Omega_3^2$ is the line element on the unit 3-sphere, written in the Hopf coordinates as
	\begin{equation}\label{eq:3sphere}
	d\Omega_3^2= d\theta^2+ \sin^2 \theta d\phi^2+ \cos^2\theta d\psi^2,
	\end{equation}
	with $\theta \in [0,\pi/2]$ and $\psi, \phi \in [0,2\pi]$. The null geodesics in a five dimensional spacetime can be written as
	\bea
	\dot t &=& h^{-1}(r), \label{eq:dott2}\\
	\dot \phi &=& \frac{\eta_\phi}{r^2  \sin^2\theta}, \label{eq:dotphi2}\\
	\dot \psi &=&\frac{\eta_\psi}{r^2  \cos^2\theta}, \label{eq:dotpsi2}\\
	\dot r^2&+&V_{eff}(r)=0, \label{eq:dotr2}\\
	r^4\dot \theta^2&=&\chi-\eta_\phi^2 /\sin^2\theta-\eta_\psi^2 /\cos^2\theta,\label{eq:dottheta2}
	\eea
	where $\chi$, $\eta_\phi$ and $\eta_\psi$ are three constants of motion. The effective potential, $V_{eff}(r)$, is given by
	\begin{equation}\label{eq:veff2}
	V_{eff}(r)=\Bigg(\frac{\chi}{r^2f(r)}-\frac{1}{f(r)h(r)}\Bigg)
	\end{equation}
	The photon sphere radius $r_p$ and the constant of motion $\chi_p$ for the unstable photon orbits could be found from the effective potential, as described previously. Once again, exploiting the spherical symmetry of the problem, we consider the null geodesics on the $\theta=\pi/2$ hypersurface only, without loss of generality. But for the $\theta=\pi/2$ hypersurface, $\eta_{\psi}=0$ (following Eq. (\ref{eq:dotpsi2}))  and hence the geodesic equations for the 5-dimensional case, Eq. (\ref{eq:dott2}-\ref{eq:dottheta2}), reduces to similar form as that for the four dimensional case. Therefore, the general expression obtained for the static shadow and the comoving shadow for the four dimensional case, Eq. (\ref{eq:genStatShdw}) and Eq. (\ref{eq:genComShdw}), are also valid for the five dimensional case. We will now apply these expressions to find the shadow of the 5D GBdS black hole.
	
	\subsection{Shadow of charged Gauss Bonnet de Sitter black holes in five spacetime dimensions}
	The spherically symmetric solution of the Einstein-Hilbert action coupled to the Maxwell field, the Gauss Bonnet term Eq. (\ref{eq:gb}) and a positive cosmological constant $\Lambda=(d-1)(d-2)H_0^2/2$ in $d=5$ dimensions is given by substituting the metric functions in Eq. (\ref{eq:genMetric5D}): $h(r)=f^{-1}(r)=h_{GB}(r)$ with the function $h_{GB}$ being given by \cite{Wiltshire:1988uq,Cai:2003gr}
	\begin{equation}\label{eq:metricfunc}
	h_{GB}=1+\frac{r^2}{4\alpha}\Bigg(1\pm\sqrt{1+8\alpha H_0^2+\frac{64M\alpha}{3\pi r^4}-\frac{2Q^2\alpha}{3\pi r^6}}\Bigg)
	\end{equation}
	\begin{figure}[h]
		\includegraphics[width=8.5cm]{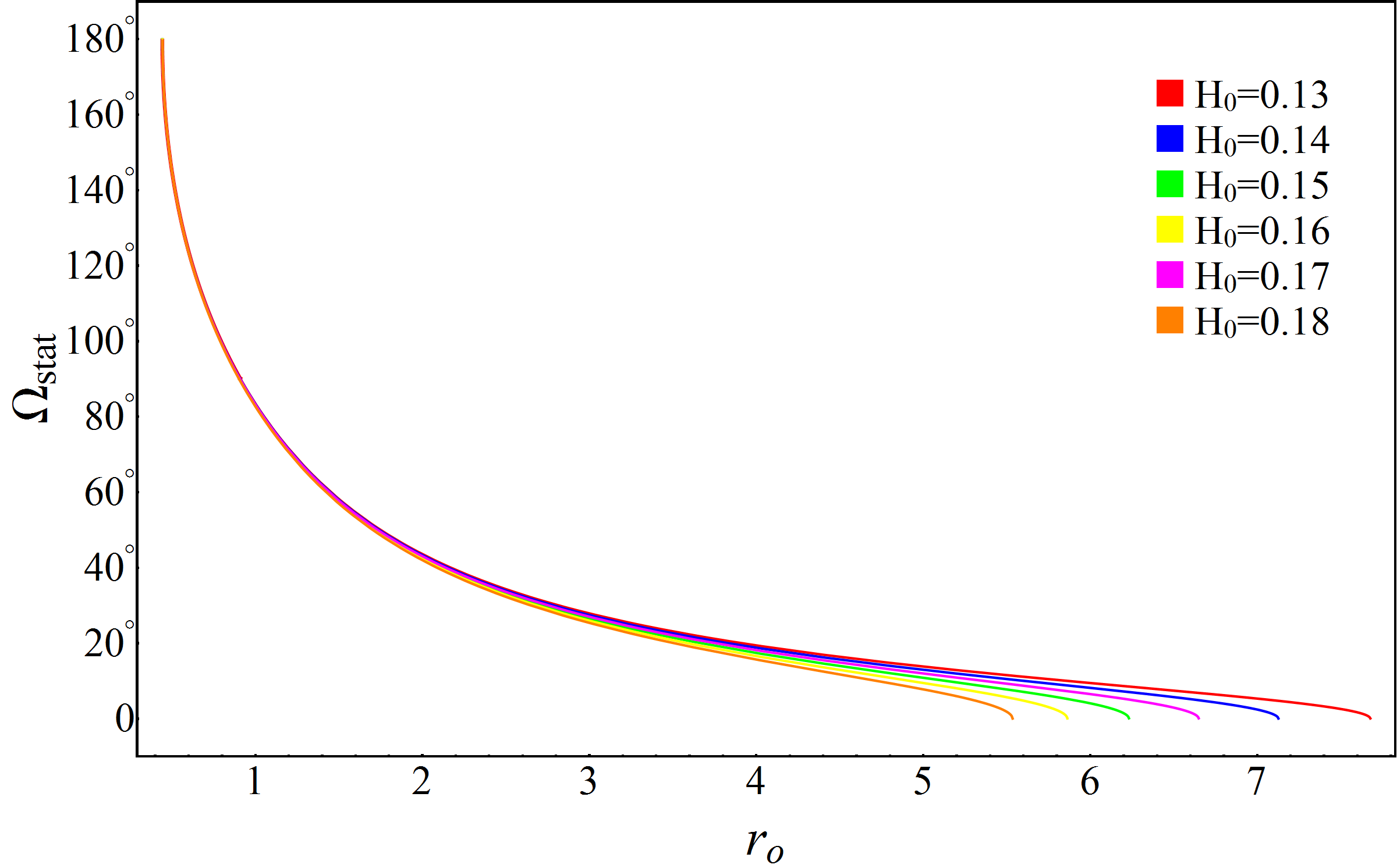}
		\centering
		\caption{The figure plots the behaviour of the static angular radius of the shadow of charged Gauss Bonnet dS black hole with $r_o$ for $M=1$, $Q=0.5$, $\alpha=0.31$ and six different values of $H_0$}
		\label{fig:GBRNdSH0stat}
	\end{figure}
	Here, $\alpha$ is the Gauss Bonnet coupling, $M$ is the mass and $Q$ is the charge of the black hole. Note that despite having a positive cosmological constant, the system has two branches, corresponding to the two signs in Eq. (\ref{eq:metricfunc}), with asymptotic dS ({negative} sign) and AdS ({positive} sign) behaviour. It has been shown that the spacetime is unstable for the branch with the positive sign \cite{deser} and for $d\geq6$ \cite{Cai:2003gr}. Hence, we are interested in the branch with {negative} sign in Eq. $(\ref{eq:metricfunc})$ and $\alpha\geq0$. Fig. \ref{fig:GBRNdSH0stat} and \ref{fig:GBRNdSH0} plots the behaviour of the static and the comoving shadow. Note that the variation in the shadow due to different values of $Q$ is insignificant and hence we don't explicitly show it here.
	\begin{figure}[h]
		\includegraphics[width=8.5cm]{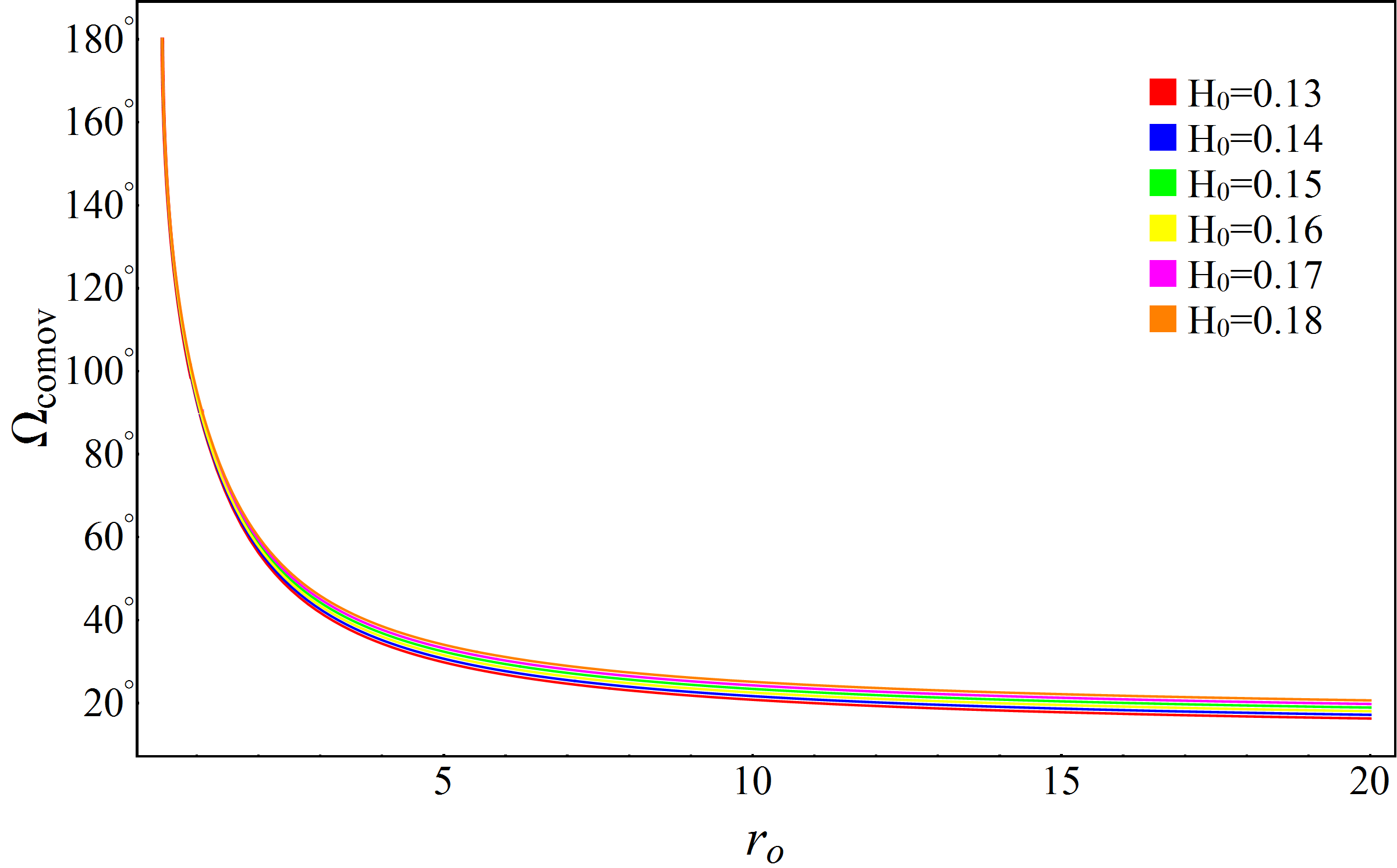}
		\centering
		\caption{The figure plots the behaviour of the comoving angular radius of the shadow of the charged Gauss Bonnet dS black hole with $r_o$ for $M=1$, $Q=0.5$, $\alpha=0.31$ and six different values of $H_0$}
		\label{fig:GBRNdSH0}
	\end{figure}
	
	{Note} that similar to the Reissner Nordstr\"{o}m de Sitter case and the four dimensional Gauss Bonnet de Sitter case, the static observers are constrained between the horizons and this feature is showcased by the termination of the static shadow at the cosmological horizon in Fig. \ref{fig:GBRNdSH0stat}, whereas the comoving shadow {approaches} a constant value as the observer tends to spatial infinity in Fig. \ref{fig:GBRNdSH0}.
	The asymptotic behaviour of the comoving and static shadow of the five dimensional charged Gauss Bonnet de Sitter spacetime could be written with a similar equation as the four dimensional case, with the effective Hubble's constant $H_{\rm eff}$ replaced by
	\begin{equation}
	H_{\rm eff}^2=\frac{1}{4\alpha}(\sqrt{1+8\alpha H_0^2}-1).
	\end{equation}
	{We notice that the asymptotic value of the comoving shadow has a smaller dependence on the cosmological constant than the 4-dimensional case, despite having pretty similar expression. Note that the coupling constant, $\alpha$, for the 5-dimensional and the 4-dimensional Gauss-Bonnet de Sitter black holes are  different. There exists a scaling of $16\pi$, as described previously, and hence the coupling constant in the four-dimensional case is greater than the five-dimensional case. This is reflected in the relatively larger variation of the asymptotic behaviour of the shadow for the 4-dimensional case as compared to the 5-dimensional case.}
	
	\subsection{The effective graviton metric and the graviton shadow}\label{sec:4}
	
	It is known that in gravity theories where higher order curvature terms are considered, the photon and the graviton have different geodesics. {This is a consequence of the fact that the causal structure of Lovelock theories (Gauss Bonnet gravity being one such example) are different from that of General Relativity  \cite{ycb,Izumi:2014loa,Papallo:2015rna, lovelock}. In higher curvature theories, the characteristic hypersurface determining the causal structure of a system are non-null surfaces. These characteristics hypersurfaces turn out to be null only with respect to a different effective metric. The `effective metric' for the gravitational degrees of freedom is different from the background metric in case of higher curvature theories  \cite{lovelock}}. Thus, it would be interesting to study the dynamics of graviton propagation in higher order gravity theories and hence define a `Graviton Shadow' following the concept of the photon shadow for black holes. A study of the graviton shadow for the Gauss Bonnet black hole in asymptotically flat spacetime is given in \cite{Mishra}. For the sake of completeness, we give a brief discussion of the same in the asymptotically de Sitter case. 
	\begin{figure}[h]
		\includegraphics[width=8.5cm]{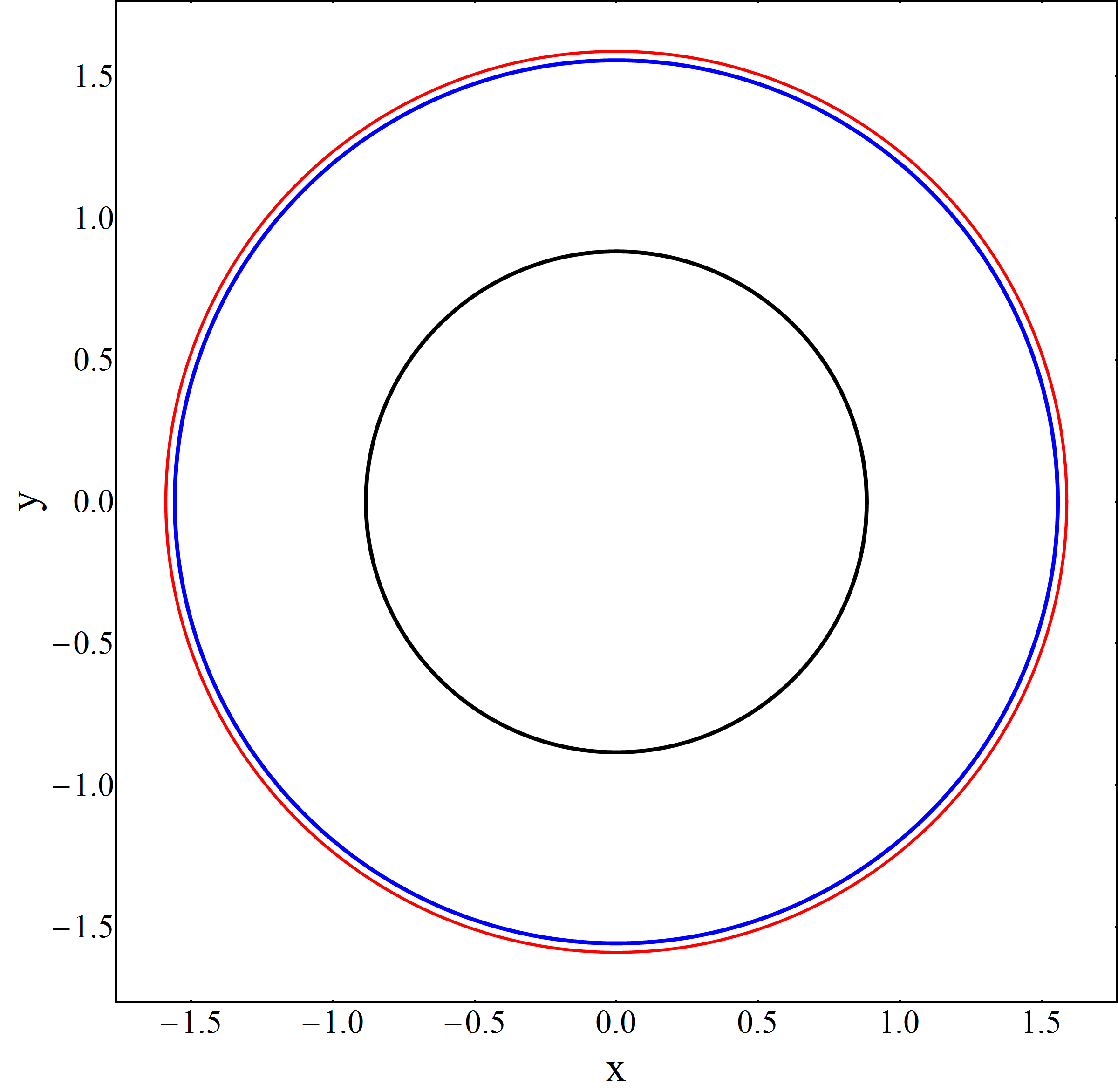}
		\centering
		\caption{The figure shows the photon shadow (red) vs the graviton shadow (blue) of a Gauss Bonnet black hole {projected on the local sky of an observer, at $r_o=3.9277$, with local coordinates $(x,y)$}. $R_p=1.58912$ and $R_g=1.55747$ with $H_0=0.15$, $\alpha=0.001$, $M=1$, $Q=0.5$. The event horizon is shown in black.}
		\label{fig:PhotonVsGravitonShadow}
	\end{figure}
	
	The effective graviton metric for the Gauss Bonnet black hole in asymptotically de Sitter space is given by \cite{lovelock}
	\begin{equation}\label{eq:gravmetric}
	ds_{eff}^2=-\bar h(r)dt^2+\bar f(r)dr^2+\bar k(r)d\Omega_3^2
	\end{equation}
	where the functions $\bar h(r)$, $\bar f(r)$ and $\bar k(r)$ are as follows 
	\bea
	\bar h(r)&=&h_{GB}(r)\Bigg(1-2\alpha \frac{h'_{GB}(r)}{r}\Bigg), \label{eq:f}\\
	\bar f(r)&=&\frac{1}{h_{GB}(r)}\Bigg(1-2\alpha \frac{h_{GB}'(r)}{r}\Bigg), \label{eq:h}\\
	\bar k(r)&=&r^2(1-2\alpha h_{GB}''(r)), \label{eq:g}
	\eea
	where a dash denotes differentiation with respect to $r$. To maintain the hyperbolicity of the metric and thus preserve causality, the terms inside parenthesis in Eq. $(\ref{eq:f})$ and Eq. $(\ref{eq:g})$ must have the same sign. Moreover, in $d=5$ dimensions, it has been found that these functions are always positive \cite{lovelock}. Next, we expand the effective potential for the radial graviton propagation in terms of $\alpha$ and keep upto the first order terms. This is because the entire problem with the exact potential is very complicated and solving the instability conditions, even using numerical approach, becomes very hard. Solving the instability conditions with the approximate potential, we plot the graviton shadow along with the photon shadow in Figure \ref{fig:PhotonVsGravitonShadow}.
	
	\section{Conclusion}
	We studied the shadow of a general spherically symmetric black hole in asymptotically de Sitter space time, i.e.  with a positive cosmological constant. It is to be stressed that our approach of computing shadows does not require the knowledge of the exact McVittie type transformation of the black hole spacetime. Another caveat in finding the comoving shadow lies in the fact that one needs to find a closed form isotropic transformation function, which, many times is not possible to get. In this work, we have shown that even without knowing the exact McVittie type transformation of the given black hole spacetime,  it is possible to calculate the comoving velocity of an observer just by using the metric functions. We have checked our results with already existing calculations \cite{Perlick} for the Schwarzschild de Sitter black hole using our approach and found an exact agreement. Thereafter, using this expression for the velocity of the comoving observer with respect to a static observer, we found the general expressions for the angular radius of the shadow with respect to the static and comoving class of observers. We also showed that this formalism is not only constrained to Einstein's general theory of relativity, but could be extended to include modified gravity theories. Three different test cases, the Reissner Nordstr\"{o}m de Sitter, the recently proposed 4 dimensional Gauss Bonnet de Sitter and the 5 dimensional charged Gauss Bonnet de Sitter black hole solutions, were studied using our formalism and we found some similar results as in \cite{Perlick} regarding the nature of the black hole shadow. 
	
	Although we constrained our study to five dimensional black holes, this formalism can in principle be extended to higher dimensions, as mentioned above. The important ingredient for the comoving shadow, the comoving velocity is found for $d$-dimensional spacetime and hence by looking at null geodesics in $d$-dimensional spacetimes and figuring out the static shadow, the comoving shadow can be found by using the general expression for comoving velocity Eq. (\ref{eq:genVel}) and the aberration formula.
	
	An interesting study for the future will be to see if our formalism can be extended to general axisymmetric cases. As noted in \cite{Li:2020drn}, comoving observers can only be defined in the far away region from the black hole in the axisymmetric case. We plan to present the results of such studies in future communications.


\end{document}